\documentclass[referee]{raa}
\usepackage{graphicx,times}
\usepackage{natbib}
\usepackage{amssymb,amsmath}
\bibpunct{(}{)}{;}{a}{}{,}

\usepackage[a4paper=true,dvipdfm=true,pagebackref=true]{hyperref}
\hypersetup{pdftitle = The title of my PDF, pdfauthor = My name, pdfsubject= The subject, pdfkeywords = keyword1 keyword2 keyword3}
\hypersetup{colorlinks = true, linkcolor = green, anchorcolor = red, citecolor = blue, filecolor = red, pagecolor = red, urlcolor = red}

\begin{document}

   \title{CCD astrometric observations of 2017 VR12, Camillo and Midas
}

   \volnopage{Vol.0 (20xx) No.0, 000--000}      
   \setcounter{page}{1}          

   \author{Zhen-Jun Zhang
      \inst{1,2,3}
   \and Yi-Gong Zhang
      \inst{1,2,3}
   \and Xiang-Ming Chen
      \inst{1,2}
   \and Jian-Cheng Wang
      \inst{1,2}
   \and Jie-Su
      \inst{1,2}
   }

   \institute{Yunnan Observatories, Chinese Academy of Sciences, Kunming 650216, China;  {\it zjzhang@ynao.ac.cn}\\
        \and
             Key Laboratory of the Structure and Evolution of Celestial Objects, Chinese Academy of Sciences, Kunming 650216, China\\
        \and
             University of Chinese Academy of Sciences, Beijing 100049, China\\
\vs\no
   {\small Received~~20xx month day; accepted~~20xx~~month day}}
\abstract{ We have observed three near-Earth objects(NEOs), 2017VR12, Camillo, and Midas during the year 2018. The observations were made by the 1-m telescope of Yunnan Observatory over 2 nights. Their precise astrometric positions are derived from 989 CCD observations. The theoretical positions of asteroids are retrieved from the Jet Propulsion Laboratory (JPL) Horizons System and Institut de M\'{e}canique C\'{e}leste et de Calcul des \'{E}ph\'{e}m\'{e}rides (IMCCE). The positions of three asteroids are measured with respect to the stars in Gaia DR2 star catalogue. For 2017 VR12, the mean (O-C) of right ascension and declination are -0.090$^{''}$ and -0.623$^{''}$  based on the ephemeris of published JPL, but the mean (O-C) are 3.122$^{''}$ and -0.636$^{''}$ based on the ephemeris of published IMCCE. The great difference in declination could be explained by several factors. (1)The degenerated CCD images caused by the high apparent motion speed of the object leads to the reduction of positioning accuracy. (2)The poor timing system may bring the system error, especially in the high speed direction. (3)The asteroid may be perturbed by the earth when it approaches the earth too closely.  The astrometric results show that the centroid centring method can reduce the dispersion of the non-Gaussian images compared with the PSF model method. For Camillo and Midas, the astrometric results are consistent based on two ephemerides. High-precision timing system, some astronomical effects and geometric distortion of CCD images should be carefully considered in the future works.
}

   \authorrunning{Z.J.Zhang \emph {et al.} }            
   \titlerunning{ CCD astrometric observations of 2017 VR12, Camillo and Midas }  

   \maketitle

%
%
\section{Introduction}           
\label{sect:intro}

The research of Near-Earth object (NEO) research plays an increasingly important role not only in solar system science but also in protecting our planetary environment and human society from the asteroid and comet hazard\citep{RN321}. The near-Earth asteroids (NEAs) are asteroids with the perihelion distance less than 1.3AU. Potentially hazardous asteroids (PHAs) are NEAs that have the minimum orbit intersection distances (MOID) with the earth are less than 0.05AU and have the absolute magnitude H $<$ 22.0 mag\citep{RN325,RN172}, Continuous optical and radar observations are needed for accurate orbit determination and physical characterisation, and the special attention to their orbit monitoring is required\citep{RN178,RN179}. The NEA population compared with all other asteroid populations have shorter periods, making it a as reliable set of dynamical reference frame representatives. Optical data together with radar data can be used in constraining NEA dynamics and possibly revealing more subtle, non-gravitational effects such as the YarKovsky effect\citep{RN178}. Several national and international observational efforts have been done to detect undiscovered NEOs and especially PHAs, to determine their orbital properties and impact probabilities, and to investigate their physical nature\citep{RN172}. Astrometric follow-up is also essential for targets of future radar observation, space mission and other observing campaigns. Gaia Follow Up Network for Solar System Objects (Gaia-Fun-SSO) was established to coordinate astrometric follow-up observations\citep{RN336}.\\
Many accurate astrometric observations are very useful for various purposes. In order to obtain accurate observations, some fundamental strategies are very important and should be complied with. For Near-Earth object, due to the high motion velocity, we require a quick readout but well time-recording CCD camera to obtain good quality images. For all of the astrometric observation, we used the  2$*$2 pixels binning mode to reduce the read-out time and readout noise. We estimated the accuracy of time-recording and its effect on the astrometry results of NEAs. A reference star catalogue is a key to derive the precise and accurate positions of NEAs, so we chose the newest and most precise star catalogue Gaia DR2\citep{prusti2016gaia,brown2018gaia} as our reference star catalogue.
We have observed three NEAs, 2017 VR12, (1981) Midas and (3752) Camillo using the Yunnan observatory 1-m telescope in March 2018. In Section 2, the details of observation are described. In section 3, we present the methods used to measure the positions of three NEOs and estimated some important errors. In section 4, we show the astrometric results and give some discussion. We also made some comparisons with the historical observations. Finally, we draw some conclusions in section 5.


\section{Observations}
\label{sect:Obs}
All the observations were made by Yunnan Observatory 1m telescope with the primary mirror of 1000mm diameter and 13300mm equivalent focal length. More details of the telescope and CCD detector are shown in Table~\ref{Specifications}. The site (\emph{i.e.} IAU code 286) is located at longitude E202.788$^{\circ}$, latitude N25.0294$^{\circ}$.  All of the observations are made by using the 2$*$2 pixels binning mode.

The Near-Earth asteroid 2017 VR12, and Midas were identified as PHAs. 2017 VR12 is a sub-kilometer asteroid with a somewhat elongated and angular shape, and a diameter of about 160 meters This V-type asteroid has a rotation period of approximate 1.5 hours. More details are described in Table~\ref{Detail}.

\begin{table}
\begin{center}
\caption[]{ Specifications of 1-m Telescope and CCD Detector.}\label{Specifications}


 \begin{tabular}{lr}
  \hline\noalign{\smallskip}
  \hline\noalign{\smallskip}
Approximate focal length  & $1330 cm $ \\ 
F-ration  & $13$               \\
Diameter of primary mirror  & $100 cm $      \\
CCD field of view &   7.1 $^{'}\times$ 7.1$^{'}$ \\
Size of CCD array & $2048\times2048$ \\
Size of pixel & $13.5\mu\times13.5\mu$ \\
Approximate angular extent per pixel & $0.21^{''} $\\
  \noalign{\smallskip}\hline
\end{tabular}
\end{center}
\end{table}

\begin{table}
\begin{center}
\caption[]{  Detail Information of the Three Asteroids.}\label{Detail}


 \begin{tabular}{lccccccc}
  \hline\noalign{\smallskip}
  Identification&Discoverer&Intersection distance&Magnitude&\multicolumn{2}{c}{Motion velocity}&Apparent radius&Phase\\
  &&(AU)&&\multicolumn{2}{c}{ ($^{''}/min$)}&$(^{''})$&$(^{\circ})$\\
  &&&&d(RA)/dt*cosD&d(DEC)/dt&&\\
  \hline\noalign{\smallskip}
2017 VR12 &-----& 0.0077& $\approx$12.5&2$\sim$4&-23$\sim$-42&0.006&53\\
3752 Camillo & E. F. Helin e.t. & 0.078&14.8&0.45&4.5&0.006&60 \\
1981 Midas  &C.T.Kowal& 0.0034 &14.3&-1.8&1.5&0.005&37\\
  \noalign{\smallskip}\hline
\end{tabular}
\end{center}
\end{table}

The observational sets of three NEAs are given in Table~\ref{Observations}. In the first night, all the observation were made with C filter. In the second night, a part of the observation were made by I filter. For each of these asteroids, the total number of observations was 240 for 2017 VR12, 547 for Camillo and 202 for Midas. The flat-field and bias images were taken at the beginning of the observation. To reduce the read-out noise and read-out time, all of the observations were made by using the 2*2 pixels binning mode.

\begin{table}
\begin{center}
\caption[]{  Observational Information for Three NEAs.}\label{Observations}


 \begin{tabular}{lcccc}
  \hline\noalign{\smallskip}
  Target&Obs-Date&Exposure Time&Filter\footnotemark[1]&No.\\
  &&(s)&&\\
  \hline\noalign{\smallskip}
2017 VR12&20180304&8&C&144 \\
         &20180305&4&C&30\\
         &20180305&4&I&66\\ 
3752 Camillo &20180304&6&C&272 \\
             &20180304&5&C&49 \\
             &20180305&8&C&179 \\
             &20180305&8&I&47 \\
1981 Midas   &20180304&6&C&96 \\
             &20180305&4&C&51 \\
             &20180305&20&C&19 \\
             &20180305&8&I&41\\
             &20180305&20&I&22 \\
             &20180305&60&I&14 \\
  \noalign{\smallskip}\hline
\end{tabular}
\end{center}
\end{table}
\footnotetext[1]{Filter "C" stands for clean, a neutral colour filter to keep the same optical path and "I" stands for infrared, effective wavelength midpoint $\lambda_{eff}=878 nm$ }
\section{Astrometric reduction and error estimation}
\label{sect:data}

In the observing periods of two days, the seeing in Yunnan Observatory is about 1.5$\sim$2.5 arcsec. All of the images for three asteroids were corrected by bias and flat-field images, then the positions were measured with the software Astrometrica {\it http://www.astrometrica.at/}.  In astrometric data reduction command, PSF fitting model method and centroid method were used to determine the center positions for the asteroids and reference stars. To improve matching and processing speed, we selected the stars brighter than 18 magnitude from the Gaia DR2 catalog as our reference stars. The reference stars are from the newest Gaia DR2 star catalogue\citep{RN342}which contains 1.7 billion star positions data, and the median uncertainty in parallax and position at the reference epoch J2015.5 is about 0.04mas for bright(G $ < $14 mag) sources, 0.1mas at G = 17 mag and 0.7mas at G = 20mag. The measurement processes refer to the previous work\citep{RN285,RN328}.

For small field CCD, we usually require no more than a linear fit. Using fits of a higher order always decreases the residuals for the reference stars, unless the variation of the quadratic and cubic terms from one image to the nest is significantly smaller than the value of these coefficients, a linear solution is probably an accurate representation of the true plate constants compared to a high order fit. Furthermore, it is noted that a reliable determination of higher orders in the plate constants is possible if there are many dozens of reference stars available for the solution ({\it http://www.astrometrica.at/}). In addition, when the number of reference stars is just enough available for the solution of a high order plate constants due to the poor quality centering of some stars, an over-fitting situation may occur, to cause a greater deviation from the true plate constants. Therefore, for the images with more than 12 reference stars, we choose the quadratic fit plate model to calibrate the CCD field, and for the rest, we choose the liner fit model.

Astronomical effects, such as the solar phase angle effect are considered. In the case of  phase correction, the phase angles and apparent radius are listed in the Table ~\ref{Detail}. According to \citep{RN343}, a solar phase angle with a light scattering in its surface causes an offset in its positions, described as Equations \ref{phase-correction}, where $i$ is the solar phase angle, $r$ is the apparent radius of the object, $Q$ is the position angle of the sub-solar point in the tangential plane and $C$ is a parameter related to the reflectance model adopted. For spherical object, the value of $C$ is about to 0.75.

\begin{equation}\label{phase-correction}
 \begin{pmatrix} -\triangle\alpha\cos\delta  \\ -\triangle\delta  \end{pmatrix}=\begin{pmatrix} Cs\sin{(i/2)}\sin Q \\ Cs\sin{(i/2)}\cos Q  \end{pmatrix}
\end{equation}
The shape of these three NEOs has a serious deviation from a perfect sphere, so the exact phase correction cannot be obtained. But we can estimate the phase corrections which are smaller than few milliarcsecond. In view of the small CCD field of view, We chose the topocentric astrometric positions to compare the observational ones with ephemeris ones. Considering the influence of atmospheric refraction and aberration, we should try to avoid the use of astrometric positions and replace by the apparent positions to obtain more accurate astrometric positions in the future work.

We also estimate the errors of time-recording. A well time-recording CCD camera is important to obtain accurate observation. The time of the telescope control system is synchronized with the GPS, but the time to control the CCD detector exposure is determined artificially. The unreliability of timing system may bring the system error, especially for the target with high apparent motion speed. We manually controlled the time error within 1 second during the observation. This error may bring great system error for the high motion speed objects, especially for the 2017VR12, more discussion will be given in the section 4.

A quick-moving asteroid has a trailing image (as seen from Fig.\ref{fig1}), we adopted the PSF model and centroid centring method to compare the effects of non-Gaussian images on the astrometric result.
\section{Results and discussion}
\label{sect:Results}
We divided the observations into several groups to compare the effects of different exposures and filters on the astrometric results. We also compared the observed positions of three asteroids using INPOP13C planetary ephemeris from IMCCE {\it http://www.imcce.fr/} and DE431 from JPL {\it http://ssd.jpl.nasa.gov/}.

\subsection{2017 VR12}

\begin{table}
\begin{center}
\caption[]{  Statistics of (O-C) Residuals for 2017 VR12,based on PSF-fit method}\label{VR12PSF}
 \begin{tabular}{lcccccccc}
  \hline\noalign{\smallskip}
  &\multicolumn{4}{c}{JPL}&\multicolumn{4}{c}{IMCCE}\\
  2017 VR12&\multicolumn{2}{c}{RA}&\multicolumn{2}{c}{DEC}&\multicolumn{2}{c}{RA}&\multicolumn{2}{c}{DEC}\\
  &$<O-C>$&SD&$<O-C>$&SD&$<O-C>$&SD&$<O-C>$&SD\\
  \hline\noalign{\smallskip}
 04-C8-1&-0.129&0.167&-0.542&0.343&3.507&0.167&-0.417&0.343\\
 04-C8-2&-0.067&0.250&-0.570&0.411&3.554&0.250&-0.449&0.411\\
 05-C4-1&-0.105&0.082&-0.645&0.216&2.551&0.082&-0.833&0.216\\
 05-I4-1&-0.080&0.063&-0.637&0.219&2.575&0.063&-0.825&0.219\\
 05-I4-2&-0.034&0.105&-0.718&0.270&2.493&0.105&-0.936&0.270\\
 05-I4-3&-0.020&0.042&-0.927&0.271&2.447&0.042&-1.158&0.271\\
 Total&-0.090&0.166&-0.623&0.330&3.122&0.547&-0.636&0.402\\
  \noalign{\smallskip}\hline
\end{tabular}
\end{center}
\end{table}
\begin{table}
\begin{center}
\caption[]{  Statistics of (O-C) Residuals for 2017 VR12, based on centroid method}\label{VR12CEN}
 \begin{tabular}{lcccccccc}
  \hline\noalign{\smallskip}
  &\multicolumn{4}{c}{JPL}&\multicolumn{4}{c}{IMCCE}\\
  2017 VR12&\multicolumn{2}{c}{RA}&\multicolumn{2}{c}{DEC}&\multicolumn{2}{c}{RA}&\multicolumn{2}{c}{DEC}\\
  &$<O-C>$&SD&$<O-C>$&SD&$<O-C>$&SD&$<O-C>$&SD\\
  \hline\noalign{\smallskip}
 04-C8-1&-0.173&0.127&-0.574&0.157&3.463&0.126&-0.449&0.157\\
 04-C8-2&-0.164&0.127&-0.598&0.181&3.457&0.127&-0.477&0.181\\
 05-C4-1&-0.156&0.077&-0.673&0.219&2.501&0.077&-0.861&0.219\\
 05-I4-1&-0.116&0.073&-0.638&0.223&2.539&0.073&-0.826&0.222\\
 05-I4-2&-0.071&0.137&-0.772&0.222&2.455&0.137&-0.989&0.222\\
 05-I4-3&-0.051&0.057&-0.856&0.347&2.416&0.057&-1.087&0.347\\
 Total&-0.124&0.112&-0.655&0.273&3.068&0.536&-0.659&0.367\\
  \noalign{\smallskip}\hline
\end{tabular}
\end{center}
\end{table}
 We give the statistics of  astrometric results for 2017 VR12 in Table~\ref{VR12PSF} and Table~\ref{VR12CEN} based on PSF model and centroid method respectively. Column 1 shows the details of observation information. For example, for `04-C8-1' and `05-I4-1', `04' indicates that the observation was made on March 04, 2018 and `05' indicates that the observation was made on March 05, 2018; `C8' indicates that the filter is C filter (i.e. Clear filter) and the exposure time is 8 second; `I4' indicates that the filter is I filter; the last number of the first column indicates the observation sequence. The following columns list the mean (O-C) and its standard deviation (SD) in right ascension and declination respectively. The `JPL' and `IMCCE' columns meaning the asteroid ephemeris are from JPL (i.e. DE431 planetary ephemeris) and IMCCE (i.e. INPOP13C planetary ephemeris), respectively. All units are in arc-second, and the reference stars are from the Gaia DR2 star catalogue.

Figure~\ref{VR12PSF1}-\ref{VR12PSF2} show the (O-C) residuals of 2017 VR12 based on the PSF method. The mean value of (O-C) in right ascension and declination are -0.090$^{''}$ and -0.623$^{''}$ based on JPL ephemeris, 3.122$^{''}$ and -0.636$^{''}$ compared with IMCCE ephemeris, respectively.Figure~\ref{VR12Cen1}-\ref{VR12Cen2} show the (O-C) residuals of 2017 VR12 based on the centroid methods. The mean value of (O-C) in right ascension and declination are -0.124$^{''}$ and -0.655$^{''}$ based on JPL ephemeris, 3.068$^{''}$ and -0.659$^{''}$ compared with IMCCE ephemeris, respectively. Our observations are more consistent with the ephemeris of JPL, especially in right ascension direction.

\begin{figure}
   \centering
   \includegraphics[width=60mm, angle=0]{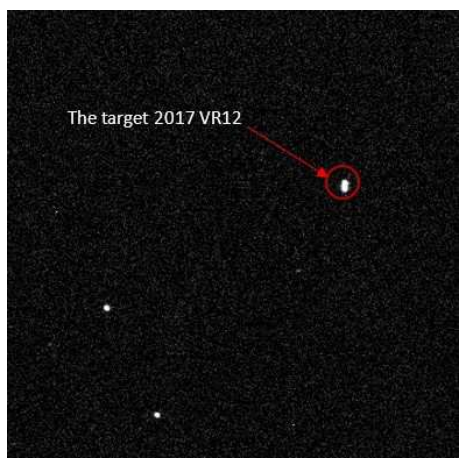}
   \caption{Part of a typical CCD images of 2017 VR12 taken on March 05, 2018 at Yunnan Observatory with 1-m telescope, the exposure time is 8 seconds. }
   \label{object1}
\end{figure}

\begin{figure}[h]
  \begin{minipage}[t]{0.495\linewidth}
  \centering
   \includegraphics[width=60mm]{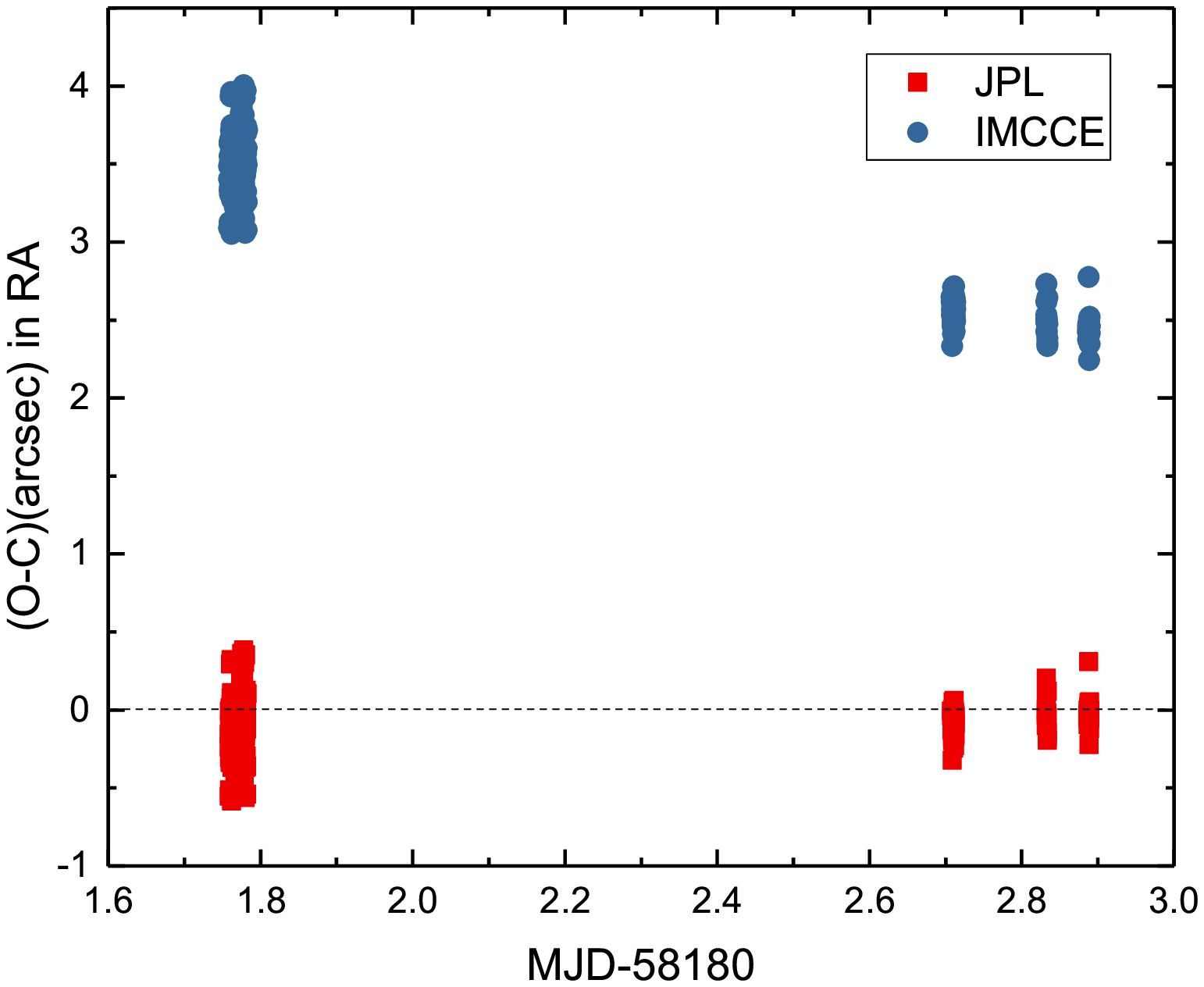}
  \end{minipage}%
  \begin{minipage}[t]{0.495\textwidth}
  \centering
   \includegraphics[width=60mm]{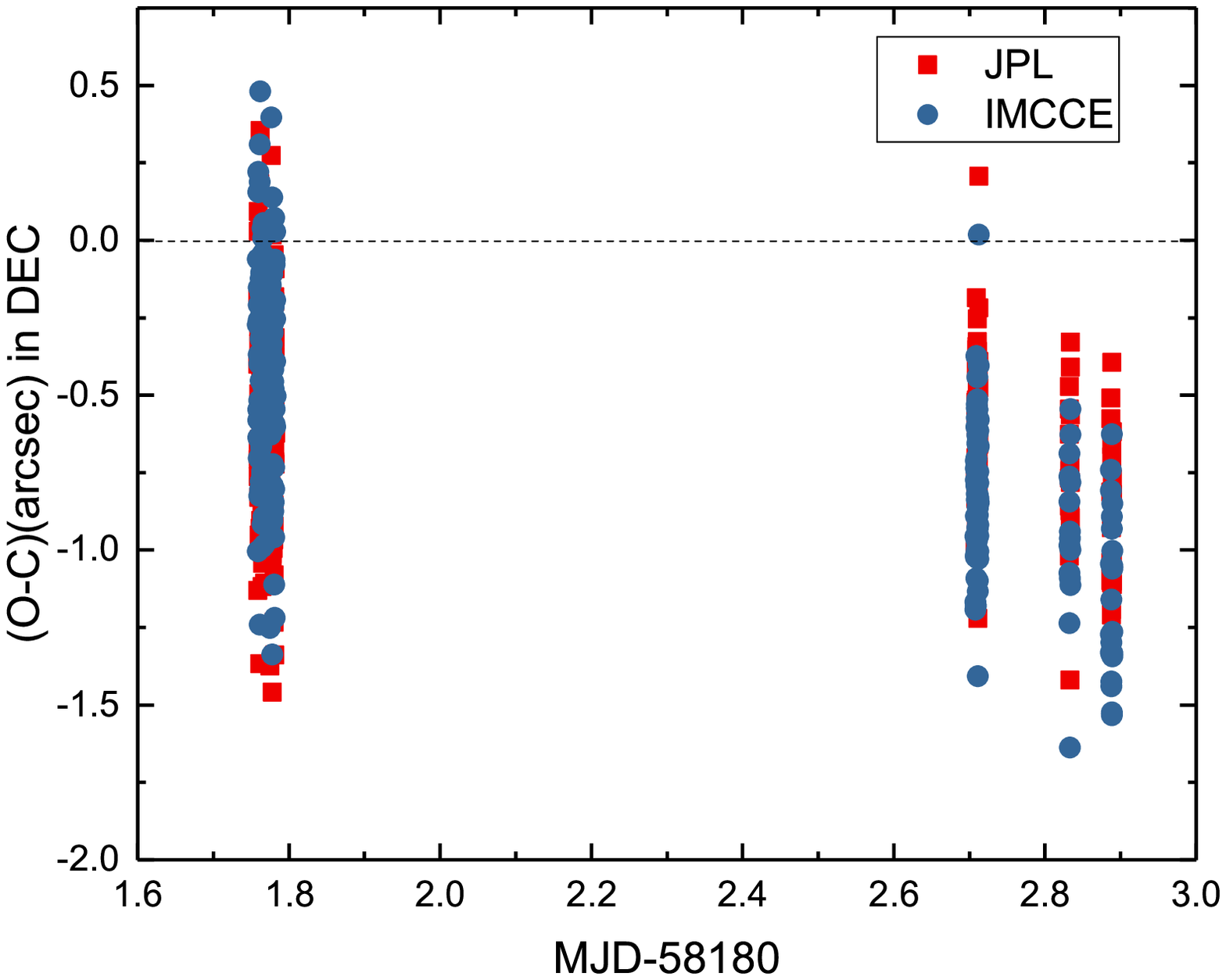}
  \end{minipage}%
  \caption{ The (O-C) residuals of the position of 2017 VR12 using different ephemerides. The red solid squares represent the (O-C) residuals using DE431 ephemeris and the blue solid circles represent the (O-C) residuals using INPOP13C ephemeris. The centering method of target adopts PSF model method.}
  \label{VR12PSF1}

\end{figure}

\begin{figure}
   \centering
   \includegraphics[width=60mm, angle=0]{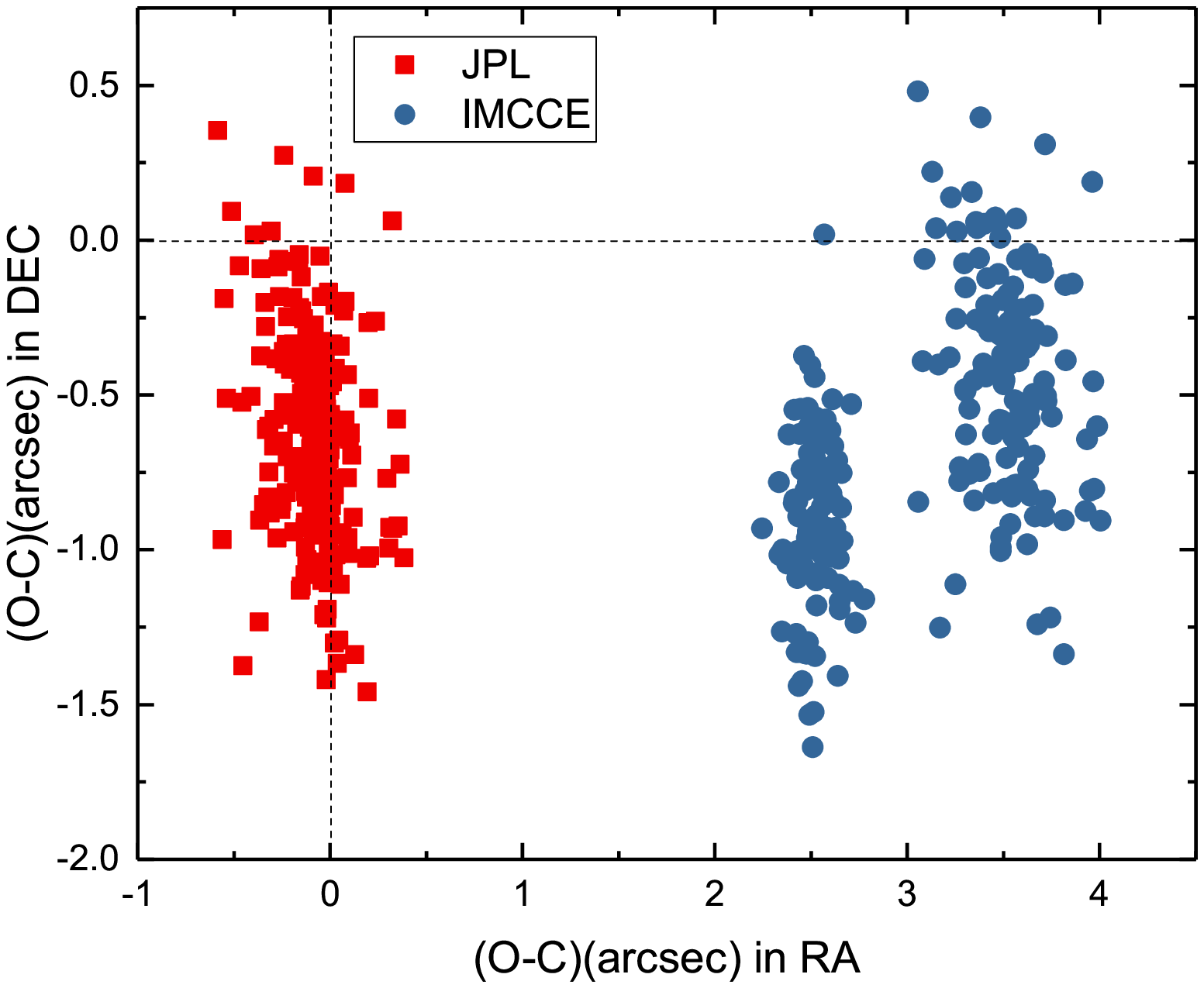}
   \caption{The (O-C) residuals of the 2017 VR12 using different ephemerides. The red points represent the (O-C) residuals using DE431 ephemeris and the blue ones represent the (O-C) residuals using INPOP13C ephemeris. The centering method of target adopts PSF model method.}
   \label{VR12PSF2}
   \end{figure}

\begin{figure}[h]
  \begin{minipage}[t]{0.495\linewidth}
  \centering
   \includegraphics[width=60mm]{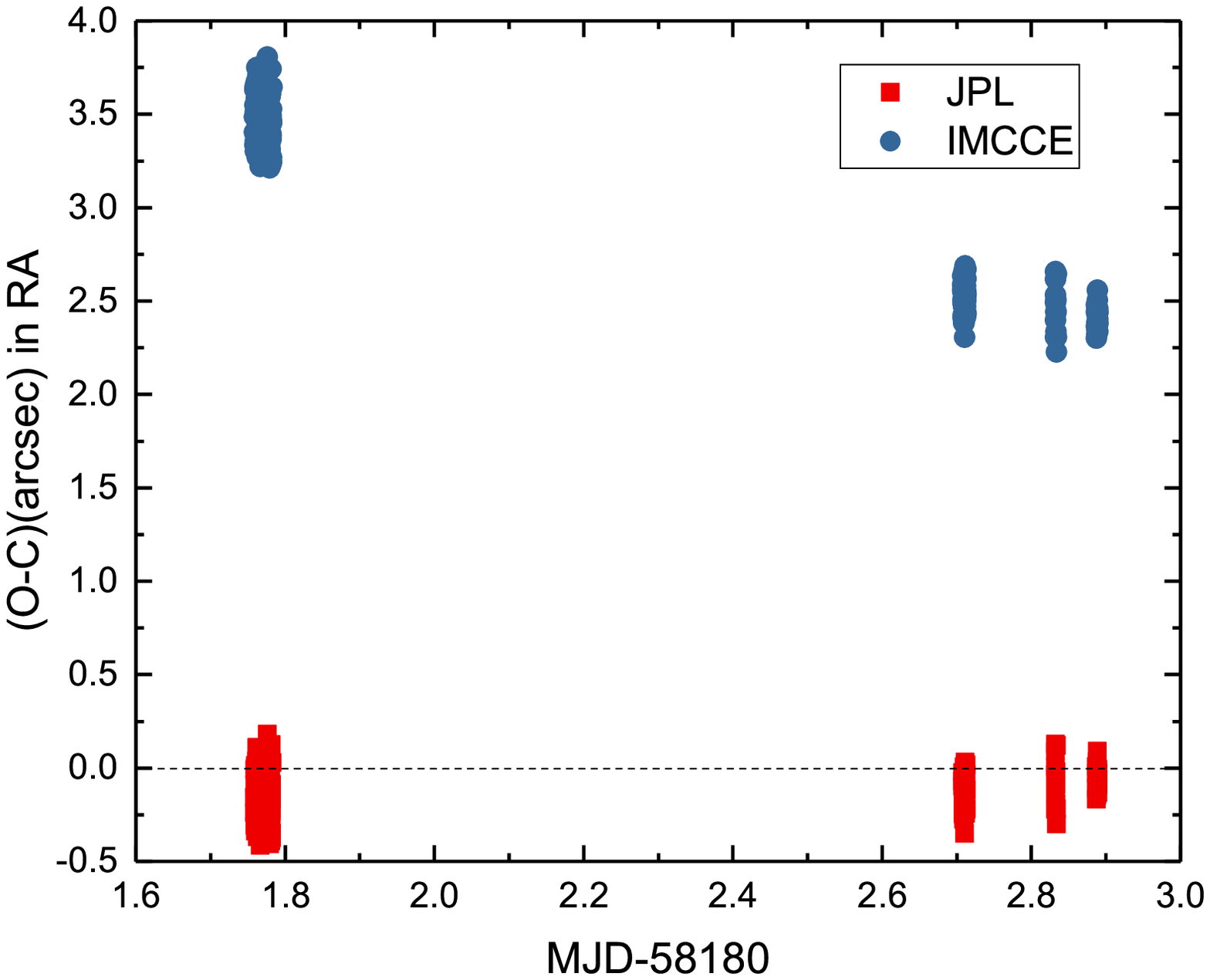}
  \end{minipage}%
  \begin{minipage}[t]{0.495\textwidth}
  \centering
   \includegraphics[width=60mm]{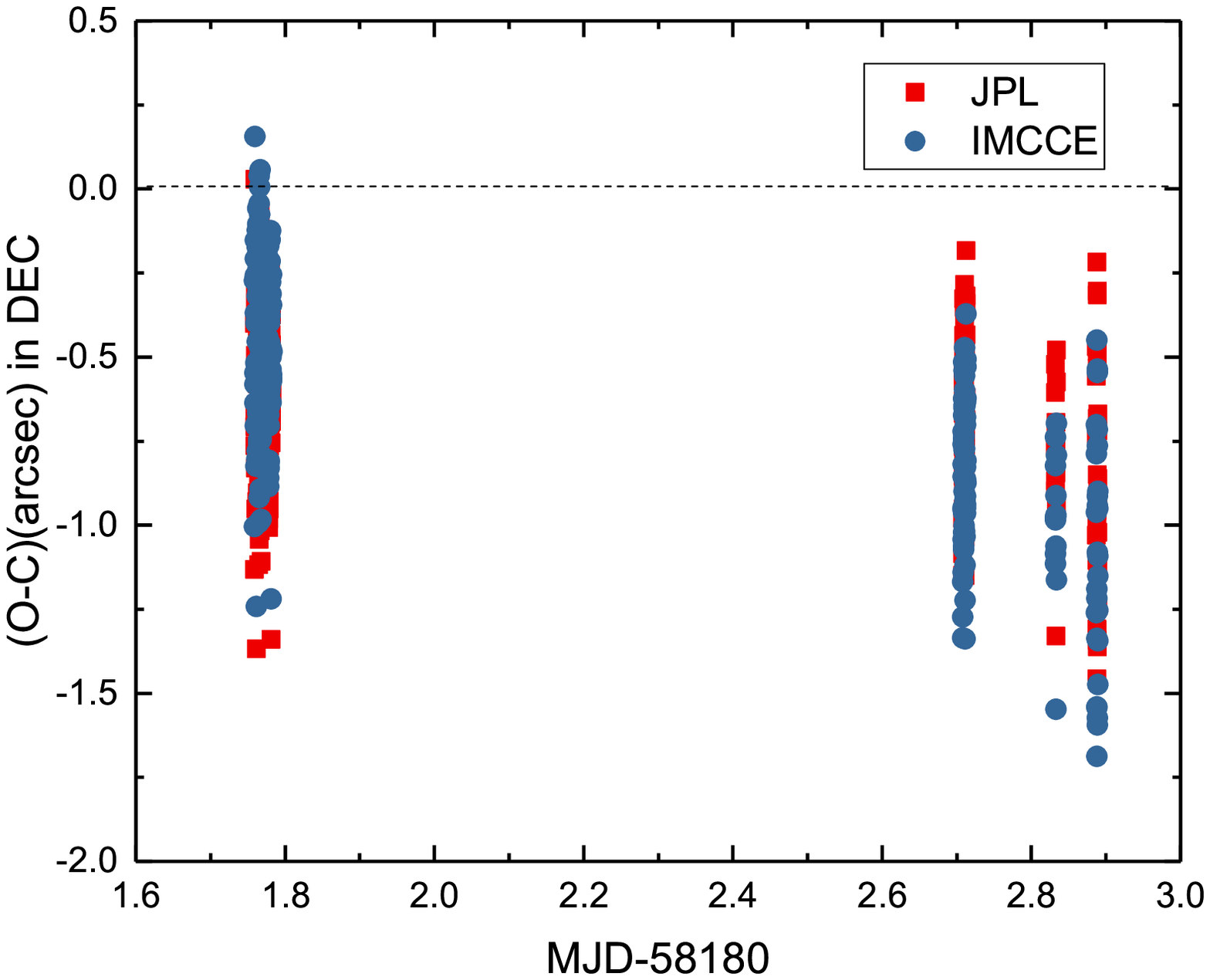}
  \end{minipage}%
  \caption{ The (O-C) residuals of the position of 2017 VR12 using different ephemerides. The red solid squares represent the (O-C) residuals using DE431 ephemeris and the blue solid circles represent the (O-C) residuals using INPOP13C ephemeris. The centering method of target adopts centroid method.}
  \label{VR12Cen1}

\end{figure}

\begin{figure}
   \centering
   \includegraphics[width=60mm, angle=0]{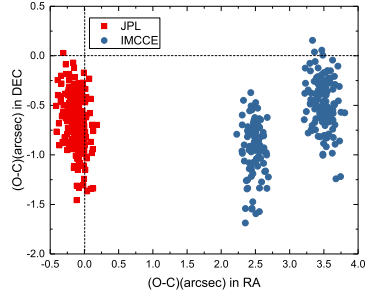}
   \caption{The (O-C) residuals of the 2017 VR12 using different ephemerides. The red points represent the (O-C) residuals using DE431 ephemeris and the blue ones represent the (O-C) residuals using INPOP13C ephemeris. The centering method of target adopts centroid method.}
   \label{VR12Cen2}
   \end{figure}

The results of 2017 VR12 are inferior to those of Camillo and Midas, one of the reasons is the high apparent motion speed. The images of 2017 VR12 were seriously distorted because of trailing, shown in Figure~\ref{object1}, especially in the declination direction. The distorted images lead to the inaccuracy of the centering, thus the dispersion of (O-C) residuals is larger. For the positioning precision, 4 seconds exposure time is better than 8 seconds. As seen in Table~\ref{Detail} the velocity component in the declination direction is larger than that in the  right ascension direction, so the dispersion in declination direction is larger. We adopted PSF model and centroid method to compare the effects of non-Gaussian images on the astrometric result. It can be seen that the astrometric results based on the centroid method have a smaller dispersion, especially when the exposure time set as 8s and the image degrades badly. But the results based on the PSF model are better agreement with JPL ephemeris. The two centering methods show significant system differences. The timing system of image acquisition should be noted, the time of the telescope control system is synchronized with the GPS, but the time to control the CCD detector exposure was determined artificially. The unreliability of timing system may bring the system error for mean (O-C), especially for quickly moving objects. Finally, the asteroid  may be perturbed by the earth when it approaches earth too closely, and the ephemeris is not very well \citep{RN183}, we can realize this from the difference between JPL and IMCCE ephemerides.


To compare our observations with other ones, we also list some history observation results of other sites({\it http://www.minorplanetcenter.net/}). Table \ref{workcompare} lists some typical residuals of the observations. It can be seen that all of the observations do not agree well with the ephemerides. The most likely reason is the lack of high-precision observational data, which limit the precision of ephemerides. The degraded image could be the main cause of the large dispersion in the declination of our work.
\begin{table}
\begin{center}
\caption[]{Compared with other observations.}\label{workcompare}
 \begin{tabular}{lccccccccc}
  \hline\noalign{\smallskip}
  &&\multicolumn{4}{c}{JPL}&\multicolumn{4}{c}{IMCCE}\\
  site&No.&\multicolumn{2}{c}{RA}&\multicolumn{2}{c}{DEC.}&\multicolumn{2}{c}{RA}&\multicolumn{2}{c}{DEC.}\\
  &&$<O-C>$&SD&$<O-C>$&SD&$<O-C>$&SD&$<O-C>$&SD\\
  \hline\noalign{\smallskip}

K28&9&-0.657&0.212&2.280&0.610&1.865&0.216&2.053&0.610\\
Z80&18&-0.115&0.276&-0.396&0.245&3.232&0.279&-0.369&0.246\\
557&28&0.605&0.288&-0.073&0.172&8.149&0.455&1.366&0.195\\
L18&9&-0.316&0.284&0.190&0.143&4.290&0.283&0.637&0.143\\
This work&240&-0.090&0.166&-0.623&0.330&3.122&0.547&-0.636&0.402\\
  \noalign{\smallskip}\hline
\end{tabular}
\end{center}
\end{table}
\subsection{Camillo}

\begin{figure}[h]
  \begin{minipage}[t]{0.495\linewidth}
  \centering
   \includegraphics[width=60mm]{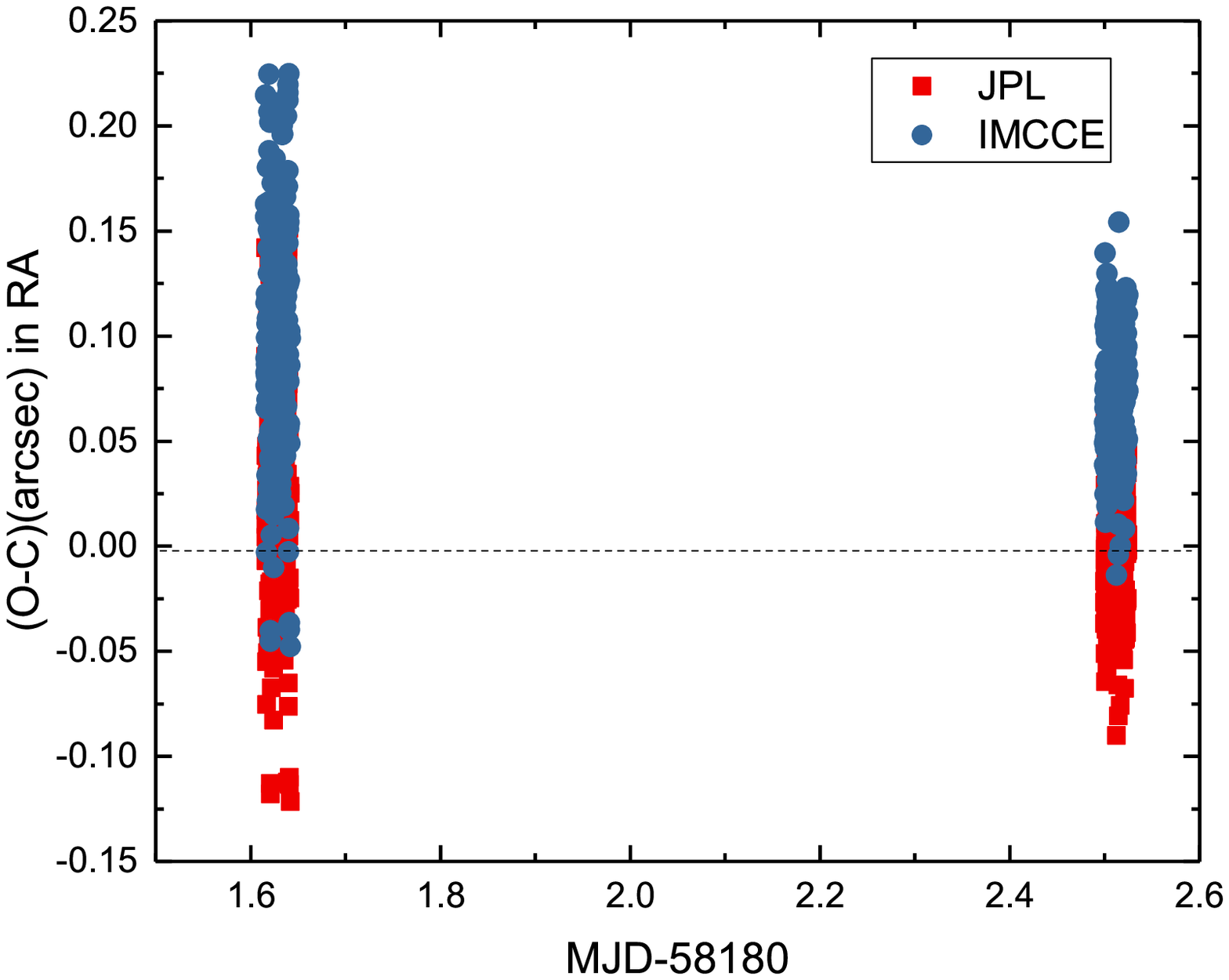}
  \end{minipage}%
  \begin{minipage}[t]{0.495\textwidth}
  \centering
   \includegraphics[width=60mm]{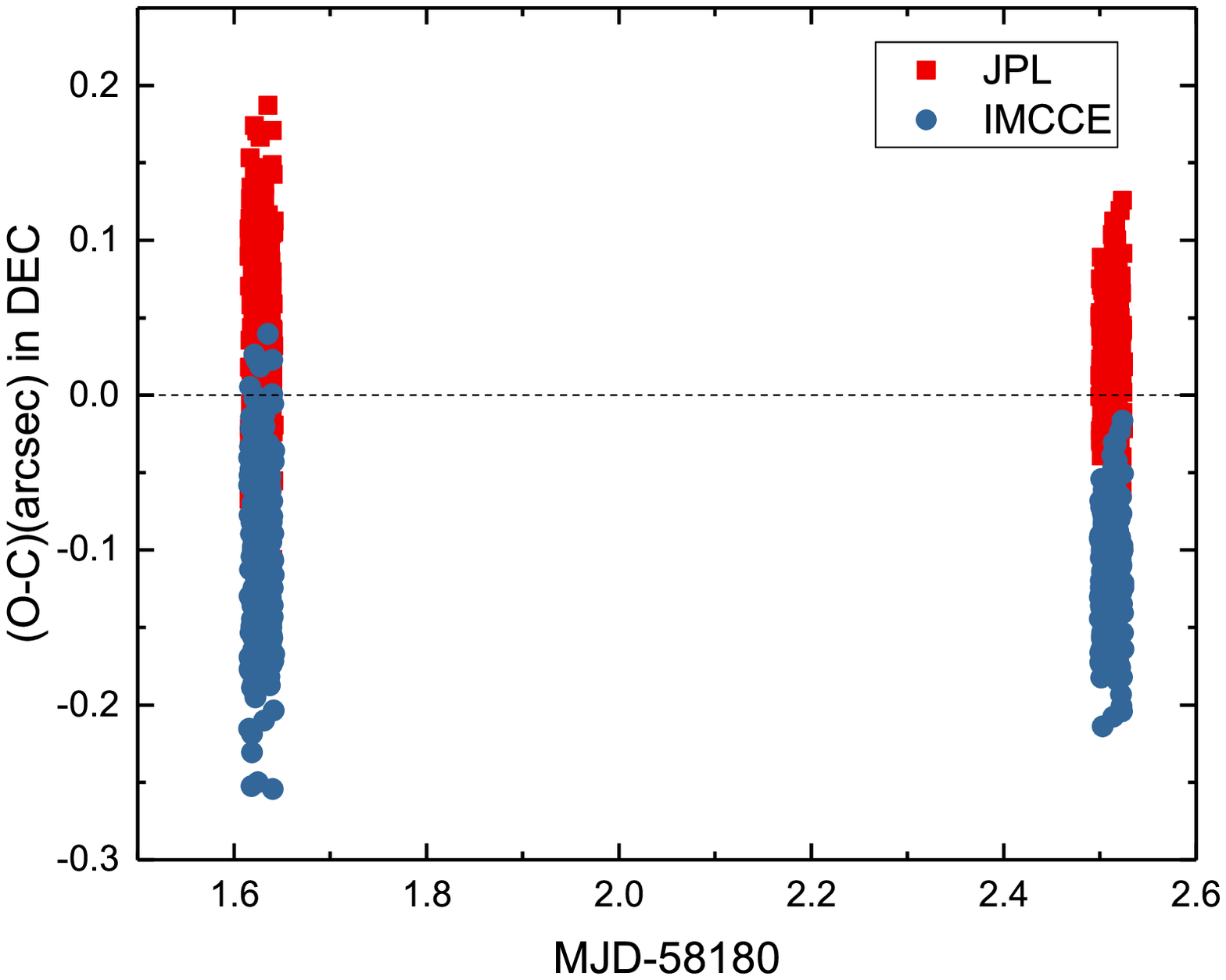}
  \end{minipage}%
  \caption{ The (O-C) residuals of the position of Camillo using different ephemerides. The red solid squares represent the (O-C) residuals using DE431 ephemeris and the blue solid circles represent the (O-C) residuals using INPOP13C ephemeris.}
  \label{Camillo1}

\end{figure}

\begin{figure}
   \centering
   \includegraphics[width=60mm, angle=0]{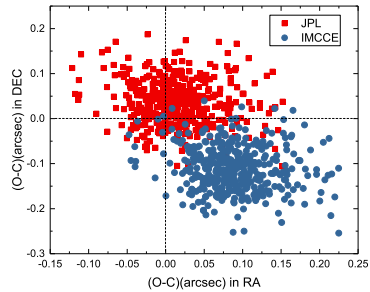}
   \caption{The (O-C) residuals of the Camillo using different ephemerides. The red points represent the (O-C) residuals using DE431 ephemeris and the blue ones represent the (O-C) residuals using INPOP13C ephemeris. }
   \label{Camillo2}
   \end{figure}

As shown in Figure~\ref{Camillo1}, Figure~\ref{Camillo2}, and Table~\ref{Camillo}, we compared the (O-C) residuals of Camillo by using different ephemerides. The mean value of (O-C) in right ascension and declination for Camillo are -0.014$^{''}$ and 0.035$^{''}$ compared with JPL ephemeris, 0.088$^{''}$ and -0.111$^{''}$ compared with IMCCE ephemeris. The dispersions of our observations are estimated to be about 0.048$^{''}$ and 0.051$^{''}$ in right ascension and declination. We can see that two ephemerides show a good agreement, especially in right ascension. It seems that the ephemeris DE431 of JPL has a better precision than the ephemeris INPOP13C in right ascension direction. In addition, the results of Camillo are better than those of 2017 VR12, one of the reasons is that Camilo’s images have sufficient reference stars. The reference stars in the Camilo’s images are more than 30 stars, but only 6-12 reference stars in the 2017VR12’s images and 6-10 reference stars in the Midas's images are used.

\begin{table}
\begin{center}
\caption[]{  Statistics of (O-C) Residuals for Camillo}\label{Camillo}


 \begin{tabular}{lcccccccc}
  \hline\noalign{\smallskip}
  &\multicolumn{4}{c}{JPL}&\multicolumn{4}{c}{IMCCE}\\
  Camillo&\multicolumn{2}{c}{RA}&\multicolumn{2}{c}{DEC}&\multicolumn{2}{c}{RA}&\multicolumn{2}{c}{DEC}\\
  &$<O-C>$&SD&$<O-C>$&SD&$<O-C>$&SD&$<O-C>$&SD\\
  \hline\noalign{\smallskip}
04-C6-1&0.028&0.051&0.044&0.058&0.101&0.051&-0.104&0.058\\
04-C6-2&0.033&0.075&0.046&0.060&0.106&0.075&-0.102&0.060\\
05-C5-1&-0.002&0.031&0.018&0.035&0.074&0.031&-0.126&0.035\\
05-C8-1&-0.007&0.029&0.025&0.042&0.069&0.029&-0.117&0.042\\
05-C8-2&0.009&0.032&0.020&0.035&0.085&0.032&-0.123&0.035\\
05-I8-1&0.005&0.035&0.043&0.048&0.081&0.035&-0.100&0.048\\
Total&0.014&0.048&0.035&0.051&0.088&0.047&-0.111&0.050\\
  \noalign{\smallskip}\hline
\end{tabular}
\end{center}
\end{table}


\subsection{Midas}

\begin{figure}[h]
  \begin{minipage}[t]{0.495\linewidth}
  \centering
   \includegraphics[width=60mm]{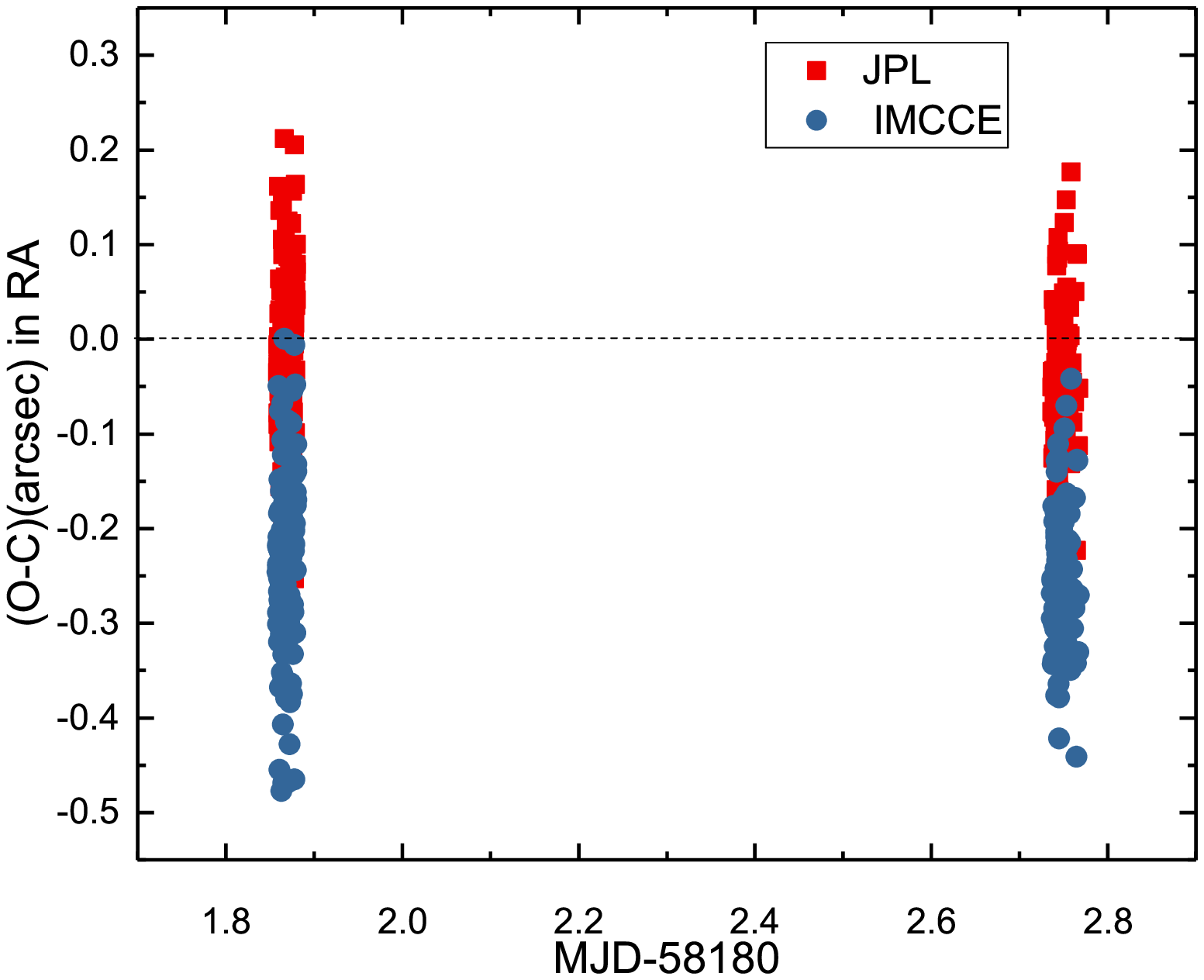}
  \end{minipage}%
  \begin{minipage}[t]{0.495\textwidth}
  \centering
   \includegraphics[width=60mm]{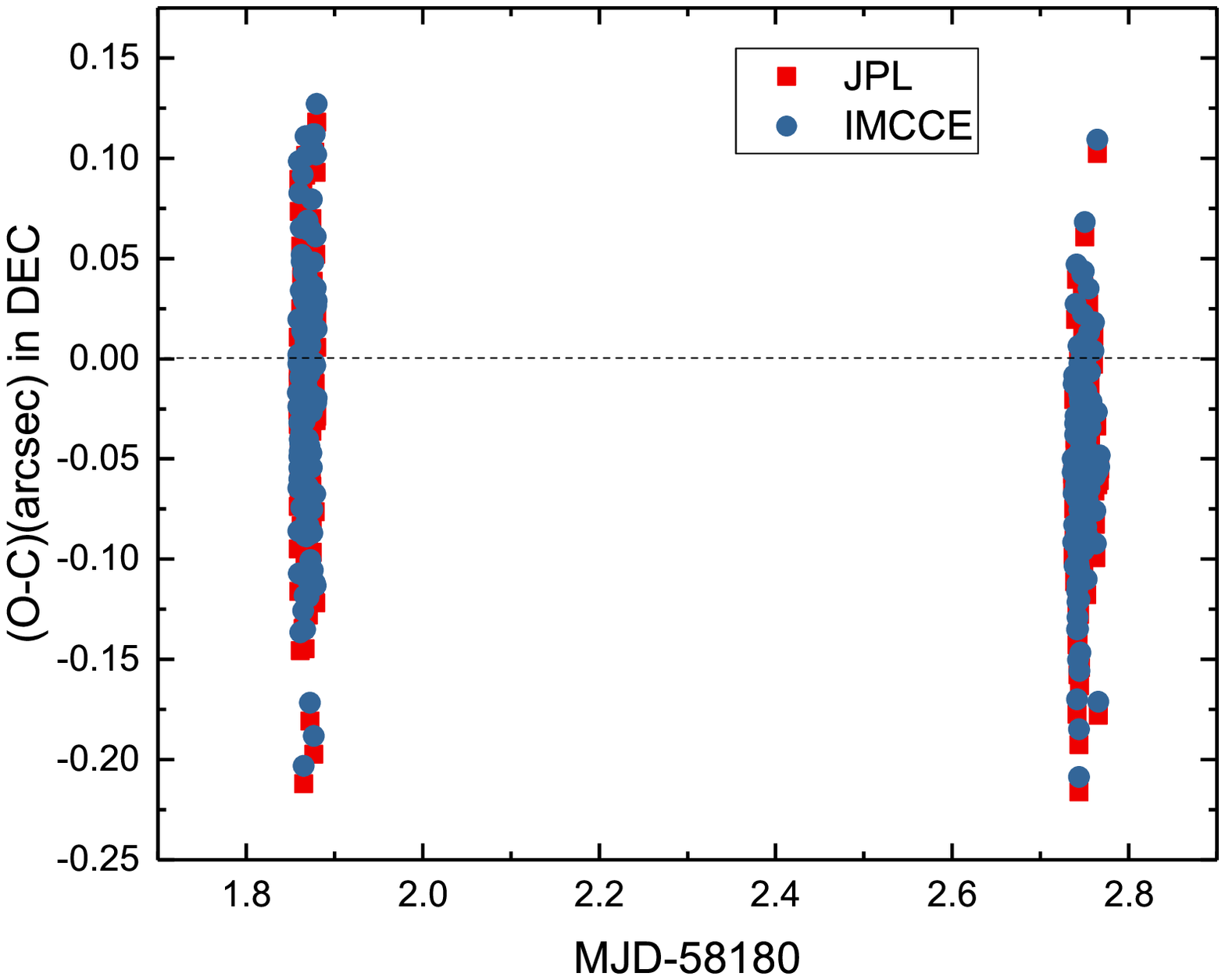}
  \end{minipage}%
  \caption{ The (O-C) residuals of the position of Midas using different ephemerides. The red solid squares represent the (O-C) residuals using DE431 ephemeris and the blue solid circles represent the (O-C) residuals using INPOP13C ephemeris.}
  \label{Midas1}

\end{figure}

\begin{figure}
   \centering
   \includegraphics[width=60mm, angle=0]{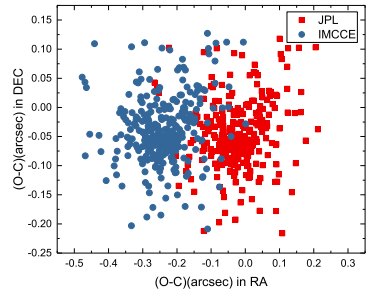}
   \caption{The (O-C) residuals of the Midas using different ephemerides. The red points represent the (O-C) residuals using DE431 ephemeris and the blue ones represent the (O-C) residuals using INPOP13C ephemeris. }
   \label{Midas2}
   \end{figure}

In Figure~\ref{Midas1}, Figure~\ref{Midas2}, and Table~\ref{Midas}, we compare the (O-C) residuals of Midas by using different ephemerides from JPL and IMCCE. The mean value of (O-C) in right ascension and declination are -0.023$^{''}$ and -0.047$^{''}$ compared with JPL ephemeris, -0.239$^{''}$ and -0.038$^{''}$ compared with IMCCE ephemeris. The dispersions of our observations are estimated to be about 0.081$^{''}$ and 0.058$^{''}$ in right ascension and declination. Due to the lack of sufficient reference stars, we extend the exposure time to 60 seconds, but the result becomes worse. Most of the brighter reference stars distribute in one direction of Midas. In order to obtain enough reference stars to solve the plate constants, sometimes we have to place the target in the corner of the image, which may bring obvious system error due to the geometric distortion of CCD field of view. \citep{RN331,RN164,RN303,RN333}. Therefore, in the next step, the geometric distortion factor should be considered to derive more accurate astrometric data.

\begin{table}
\begin{center}
\caption[]{  Statistics of (O-C) Residuals for Midas}\label{Midas}


 \begin{tabular}{lcccccccc}
  \hline\noalign{\smallskip}
  &\multicolumn{4}{c}{JPL}&\multicolumn{4}{c}{IMCCE}\\
  Midas&\multicolumn{2}{c}{RA}&\multicolumn{2}{c}{DEC}&\multicolumn{2}{c}{RA}&\multicolumn{2}{c}{DEC}\\
  &$<O-C>$&SD&$<O-C>$&SD&$<O-C>$&SD&$<O-C>$&SD\\
  \hline\noalign{\smallskip}
 04-C6-1&-0.006&0.119&-0.028&0.065&-0.217&0.119&-0.018&0.065\\
 05-C4-1&-0.023&0.061&-0.087&0.046&-0.241&0.061&-0.080&0.046\\
 05-C20-1&-0.059&0.053&-0.055&0.039&-0.277&0.053&-0.047&0.039 \\
 05-I8-1&-0.034&0.054&-0.042&0.039&-0.252&0.054&-0.035&0.039\\
05-I20-1&-0.032&0.043&-0.053&0.033&-0.250&0.043&-0.046&0.033\\
05-I60-1&-0.037&0.106&-0.048&0.062&-0.255&0.106&-0.041&0.062\\
Total&-0.023&0.081&-0.047&0.058&-0.239&0.080&-0.038&0.058\\
  \noalign{\smallskip}\hline
\end{tabular}
\end{center}
\end{table}

\subsection{The estimation of the time-recording error}
\label{TimeEstimated}
\begin{figure}
   \centering
   \includegraphics[width=\textwidth, angle=0]{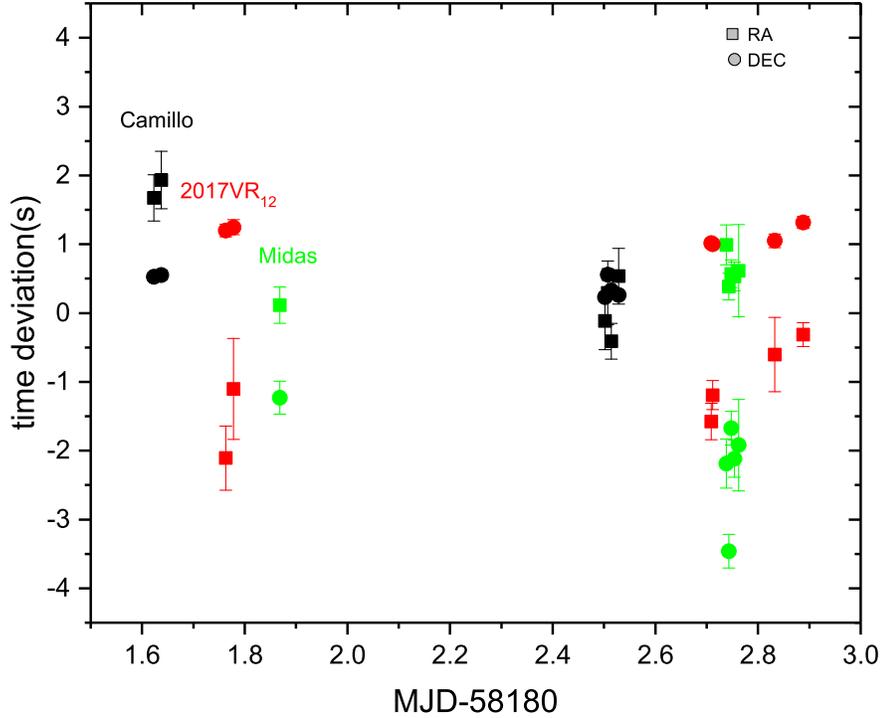}
   \caption{Time deviation estimation based on the astrometric result.The squares represent the time deviation based on the astrometric results in the right ascension and the circles represent the time deviation based on the astrometric result in the declination. The black points come from Camillo, and the red ones come from 2017VR12 , and the green ones come from Midas.}
   \label{timeerror}
\end{figure}

A well time-recording CCD camera is important to obtain accurate observation, especially for fast moving objects. As the target 2017VR12, if the timing system has 1 second system error, it will bring $0.7^{''}$ system error to the utmost in the declination direction in this observational campaign. We made some attempts to estimate the time-recording error from astrometric observation results. Now we consider this problem reversely, e. g., we infer the possible time deviation from the (O-C) results. The possible time deviation($TD$) can be obtained from  $TD = \frac{<O-C>}{v}$, where $v$ is the motion velocity of the target in right ascension or declination. The results are shown in the Figure~\ref{timeerror}. The points on the left show the results on the first day, and on the right show the ones on the second day. Two day's time are set separately and have no relevance. The error-bar are set as $\frac{SD}{v\sqrt{m}}$, where SD is the standard deviation of $<O-C>$, and $m$ is the number of observation.

As seen in Figure~\ref{timeerror}, the time deviations have no consistency in right ascension and declination, especially for 2017VR12 and midas. The difference could mainly be from the astrometric reduction and the accuracy of ephemeris, rather than the time-recording system. As seen in Table~\ref{Detail}, the target 2017VR12 has the fastest motion speed in the declination direction(red circles in Figure~\ref{timeerror}), but it has good consistency, and roughly equals to 1.2 second.
\section{Conclusions}
\label{sect:conclusion}
We have presented the results of our astrometric CCD observations of three asteroids using the 1-m telescope at Yunnan Observatory. During the reduction, the Gaia DR2 star catalogue was used to match the stars in the field of view. We have used the ASTROMETRICA software tool for astrometric data reduction of CCD images. The ephemerides of INPOP13C from IMCCE and DE431 from JPL show inconsistence for the asteroid 2017 VR12. We find that the difference between two ephemerides is about 3$^{''}$ in the right ascension direction for the asteroid 2017VR 12. We adopted two centering methods, and found that the centroid method can conspicuously reduce the dispersion of the non-Gaussian images compared with the PSF model method. The observations of Camillo and Midas are consistent based on two ephemerides,especially for Camillo, the mean (O-C) residuals and standard deviations are under 0.05$^{''}$. To derive more accurate astrometric data of Near-Earth objects, especially fast moving objects, we should use the precise timing system during the observation, and consider the geometric distortion of CCD images in the processing of astrometric positions. In addition, some astrometrical effects should be considered carefully.

\normalem
\begin{acknowledgements}
We acknowledge the support of the staff of the 1-m telescope at Yunnan Observatory. This research work is financially supported by the National Nature Science Foundation of China(grant nos.11503083,nos.11403101).This work has made use of data from the European Space Agency (ESA) mission
{\it Gaia} (\url{https://www.cosmos.esa.int/gaia}), processed by the {\it Gaia}
Data Processing and Analysis Consortium (DPAC,
\url{https://www.cosmos.esa.int/web/gaia/dpac/consortium}). Funding for the DPAC
has been provided by national institutions, in particular the institutions
participating in the {\it Gaia} Multilateral Agreement.
\end{acknowledgements}

\bibliographystyle{raa}
\bibliography{bibtex}

\end{document}